\documentclass[a4paper,10pt,fleqn]{article}
\usepackage{amsmath}
\usepackage{amssymb}
\usepackage{type1cm}
\usepackage{amsthm}
\usepackage[dvipdfmx]{graphicx}
\usepackage{fancybox,ascmac}
\usepackage{empheq}
\usepackage[hyphens]{url}
\usepackage{here}
\usepackage{fancyhdr}
\usepackage{bm}
\usepackage{color}
\usepackage{indentfirst}
\usepackage[hang,small,bf]{caption}
\usepackage[subrefformat=parens]{subcaption}
\usepackage{lineno}
\usepackage[super]{cite}
\theoremstyle{plain}

\title{\Large
\textbf{
The Mechanism of Pattern Transitions between Formation and Dispersion
}}
\author{Shin NISHIHARA\thanks{
Graduate School of Mathematics, Nagoya University. Furocho, Chikusaku, Nagoya, 464-8602, JAPAN. E-mail: shin.kurokawa.c8@math.nagoya-u.ac.jp}, and Toru OHIRA\thanks{Graduate School of Mathematics, Nagoya University}}
\date{}
\begin{document}
\maketitle
\section*{Abstract}
\textbf{
The patterns observed on the body surface of living organisms have traditionally been attributed solely to ecological strategies. However, this study investigates a fascinating phenomenon in \textit{Pelodiscus sinensis}, where patterns formed on the plastron during embryonic and juvenile stages, which are not externally visible, ultimately disappear in adulthood. This exploration suggests the existence of mechanisms beyond ecological purposes in the formation of body surface patterns in living organisms.
\\
\indent
While numerous studies have examined pattern formation mechanisms \cite{bib58, bib59}, limited research has focused on the dispersion of preexisting patterns. This study aims to investigate the actual dispersion of patterns on the plastron of \textit{P.sinensis}. During embryonic stages, distinct black patterns gradually emerge on the vivid orange plastron. However, as the organism undergoes significant growth, these patterns disperse, resulting in a whitened plastron. Our research explores the role of osteoblasts expressing the enzyme \textit{cyp26b1}, retinoic acid, and melanoblasts/melanophores on the ventral part of the plastron, which undergoes a morphological transition from a narrow and rod-shaped structure in juveniles to a broader and flatter structure in adults \cite{bib46, bib61}.
\\
\indent
To understand the changing patterns and coloration of the plastron, we propose a hypothesis based on a reaction-diffusion system with a time-dependent growing spatial domain. This mathematical framework suggests the occurrence of the dispersion phenomenon. Specifically, we focus on the dilution term within the system under the growing-domain condition. Previous studies have investigated growing-domain effects \cite{bib45, bib48, bib56, bib62, bib66, bib67}, but our study specifically addresses the role of the dilution term. In the context of black-pattern formation, we propose that variations in retinoic acid concentration, indirectly influenced by osteoblasts expressing the enzyme \textit{cyp26b1} during the embryonic/juvenile stage, contribute to the observed patterns \cite{bib63, bib64, bib68}. This hypothesis is grounded in the concept of prioritized osteogenesis and ossification during the embryonic/juvenile phase.
\\
\indent
This study sheds light on the intricate mechanisms underlying pattern dispersion and formation on the plastron of \textit{P.sinensis}. It expands our understanding of the species' survival strategy, highlighting the significance of bone biology and retinoic acid regulation. The findings have broader implications beyond \textit{P.sinensis}, contributing to our knowledge of pattern formation and bone development in other organisms.
}
\section{Introduction}
The origin and function of patterns observed on the body surface of living organisms have long been a topic of interest, prompting questions regarding their formation solely based on ecological strategies. For instance, zebras in the African grasslands exhibit white and black striped patterns, which have been speculated to serve as camouflage (in fact the white patterns appearing on a black background). Similarly, giant pandas possess black patterns that may aid in thermal insulation of their body parts in China's cold environment. It is evident that both species have relied on their morphology, specifically the presence of black patterns, for survival. If patterns were to disappear and yet no changes occurred in the ecology of certain species, it would imply that they were not solely formed for ecological purposes. This raises the possibility that the black patterns observed on the plastron of the living organism \textit{P.sinensis} during its embryonic and juvenile stages, which eventually vanish in adulthood, may serve a purpose beyond ecological considerations. Two perplexing questions arise: why were the black patterns formed on the inconspicuous ventral side of the organism, and why do the patterns disperse once they have been established? By elucidating the mechanisms underlying this phenomenon in \textit{P.sinensis}, we may uncover alternative mechanisms for pattern formations, including the black patterns, on the body surface of living organisms that extend beyond ecological purposes.
\\
\indent
To date, despite numerous studies on pattern formation, there is a notable gap in our understanding of the dispersion of previously-formed patterns. One promising mechanism is the concept of a growing domain, where patterns are robustly formed as the domain expands over time \cite{bib60}. However, this study proposes a distinct mechanism observed in the case of \textit{P.sinensis}, specifically the robust dispersion of patterns previously expressed during the embryonic stages \cite{bib65} on the plastron. This phenomenon warrants investigation from a morphological perspective. It is particularly intriguing to consider the rationale behind the waste of previously-formed patterns during the organism's growth, especially given the vulnerable nature of juvenile \textit{P.sinensis}, which must evade potential predators after carefully concealing itself. Remarkably, the plastron of each juvenile displays vibrant orange coloration, with contrasting black patterns, despite the dispersion of patterns and the eventual whitening of the plastron in adult \textit{P.sinensis}. This raises questions about the effectiveness of this strategy for survival. Consequently, our hypothesis posits that the juvenile undergoes two crucial stages of robust and rapid growth for its survival strategy: (i) the embryonic stage and (ii) the growing-domain stage. The former stage leads to the expression of black patterns on the vivid orange plastron, while the latter stage results in pattern dispersion and the uniform whitening of the orange plastron, despite the melanophores not experiencing significant depletion \cite{bib68}.
\section{Modeling of Pattern Dispersion}
During the embryonic stages of \textit{P.sinensis}, the plastron begins to exhibit pigmentation, with the emergence of black spots \cite{bib65}. Notably, a previous study \cite{bib64} highlights the significance of dietary and circulating $\beta$-carotene in turtle pigmentation, as demonstrated through a controlled experiment comparing plasma levels of $\beta$-carotene in turtles provided with and deprived of carotenoids \cite{bib64}. These findings suggest that the juvenile \textit{P.sinensis} receives carotenoid supplementation from its yolk during morphogenesis \cite{bib65}. Building upon this knowledge, the present study puts forth hypothetical mechanisms to explain the formation, dispersion, and subsequent whitening of the black patterns on the vibrant orange plastron as follows:
\begin{figure}[H]
\begin{center}
\includegraphics[scale=0.30]{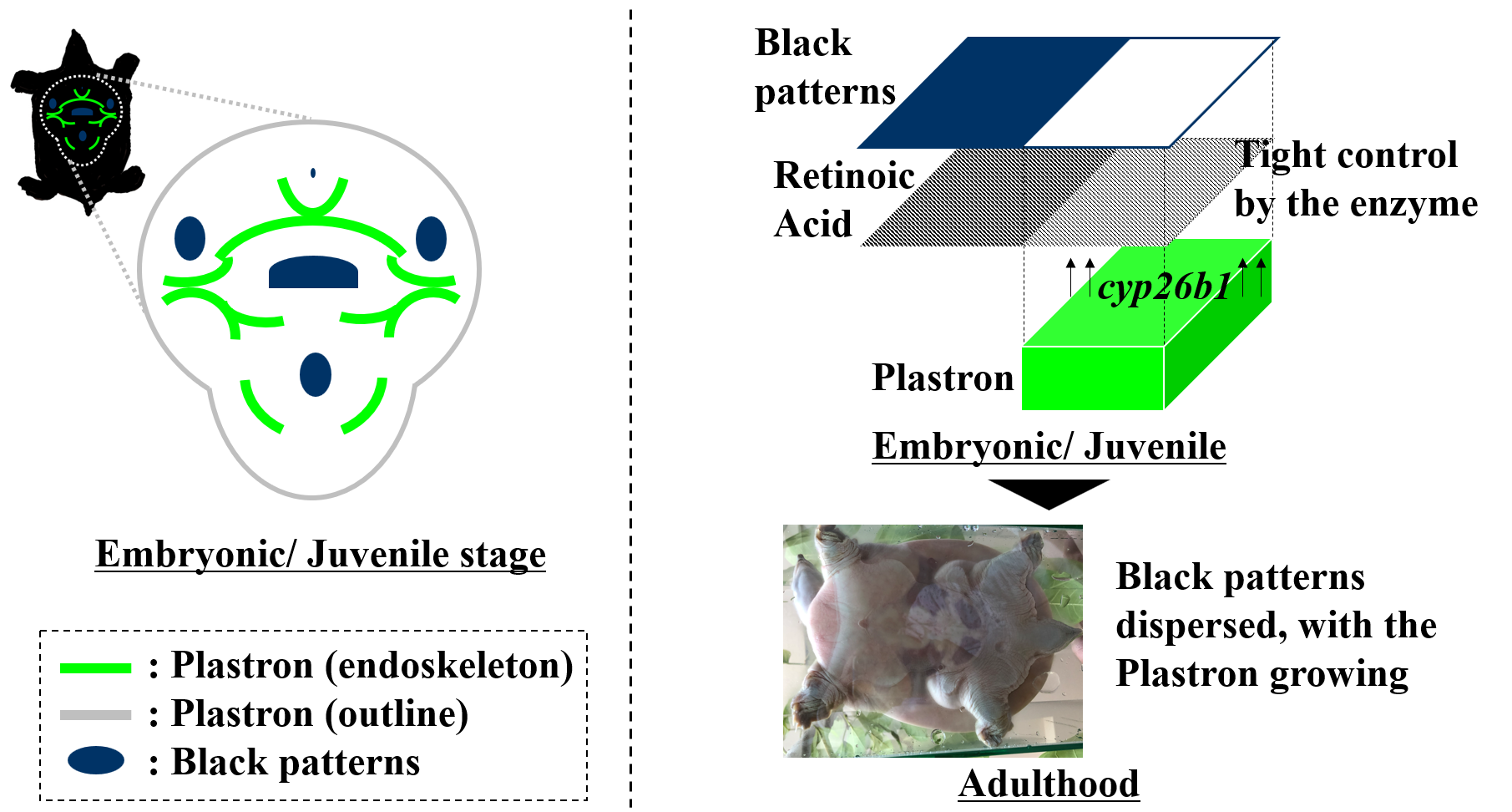}
\end{center}
\caption{The plastron and the black patterns are formed during \textit{P.sinensis} embryonic stages (\textit{left}), and the concentration gradient of retinoic acid, tightly controlled by the enzyme \textit{cyp26b1} within osteoblasts in the vicinity of the plastron, is generated between the tightly controlled domain and the uncontrolled, which leads to a relative increase in the number of melanoblasts/ melanophores (\textit{top right}). In adulthood, as $P.sinensis$ grows, the black patterns disperse with the plastron whitened (\textit{bottom right}).\label{Fig-1}}
\end{figure}
\begin{enumerate}
\item During the embryonic stages, the survival strategy of \textit{P.sinensis} prioritizes osteogenesis and ossification processes. As part of this strategy, the enzyme \textit{cyp26b1} is expressed within osteoblast cells to tightly regulate the concentration of retinoic acids (referred to as $A[M]$) \cite{bib63}. This regulatory mechanism is supported by a study \cite{bib69} demonstrating that high concentrations of vitamin A (represented by retinoic acids in this study) inhibit bone differentiation, while low concentrations promote osteoblastic activity. Throughout the embryonic stages ($T < T_g$), $A$ is strictly controlled by a specific threshold, $A_{th}[M]$, in the vicinity of bones (represented by the domain $\Omega^c$) \cite{bib46}. This control mechanism ensures that the concentration within the domain $\Omega$ is unaffected by the tight regulation facilitated by the enzyme \cite{bib63}.
\item The concentration of cells ($C[cell/mL]$) that express enzymes responsible for converting carotenoids to retinoic acids \cite{bib51, bib53} exhibits a logistic increase across the entire domain, $\Omega\cup\Omega^c$, reaching a certain carrying capacity ($K_C[cell/mL]$). As $C$ increases, $A$ also shows a logistic increase\footnote{The same results are led by a Malthus increase for $a$, i.e., $\gamma a \hspace{1mm}c(1-c)$, in the second equation for $a$ in the system ($\ref{Eq-06}$) based on numerical solutions.}, reaching a specific carrying capacity ($K_A[M]$), provided that the precursor molecules (carotenoids) are not depleted. Due to the existence of mechanisms that avoid excessive retinoic acid concentrations for osteogenesis or ossification \cite{bib54, bib63, bib68}, it is considered that the increase of $A$ is proportional to the logistic increase of $C$, which suppresses the excessive retinoic acid concentrations, rather than simply being proportional to the linear proliferation of $C$. Based on this understanding, our hypothesis in this study is that the precursors, carotenoids, are not depleted, which necessitates a more stringent condition for pattern dispersion to occur.
\item The increase of $A$ triggers an increase in melanoblasts (equivalent to melanophores), leading to an elevation in the concentration of melanin ($M[M]$) due to the upregulation of genes involved in melanin synthesis \cite{bib68}. Once $A$ surpasses a specific threshold ($A_M[M]$), melanoblast differentiation (equivalent to melanin production) commences \cite{bib59} in a logistic manner, reaching a certain carrying capacity ($K_M[M]$). In this study, we propose an additional stringent condition for pattern dispersion, which does not require the metabolism of melanin.
\item As $C$ approaches $K_C$, the domain undergoes significant growth during the growing-domain stages ($T\geq T_g$). This growth is attributed to the multiplication of cells, as it is generally observed that organisms increase in size as their cells proliferate. Simultaneously, the tight control exerted by the enzyme \textit{cyp26b1} is released \cite{bib63}, and an unknown threshold $A^*_{th}$ is introduced across the entire domain ($\Omega\cup\Omega^c$), with zero-flux boundary conditions set on the boundary ($\partial(\Omega\cup\Omega^c)$). In this study, we propose an additional stringent hypothetical condition for pattern dispersion that does not require the removal of tight control.
\item Since it is considered that the strain rate for the linear growth of the entire domain depends on $C$, a certain function, $\mathcal{R}(C)$, is defined below such that the strain rate is smaller when $C$ is away from $K_C$ and is approximately equal to $R[1/sec]$ when $C$ is closer to $K_C$:
\begin{equation}
\label{Eq-01}
\mathcal{R}(C) := \frac{R\hspace{1mm}h}{h+\Big( C/K_C -1\Big)^2}\hspace{1mm},
\end{equation}
where $h$ ($h>0$) is a dimensionless hatching coefficient which represents the degree of ease to hatch, meaning that a lower value, for instance, indicates a sufficient development of the fundamental body structure inside an egg prior to hatching.
\end{enumerate}

Mathematically, the two-dimensional spatial variables on the growing domain are described as follows\footnote{see \ref{App-01} for intermediate formulae}:
\begin{equation}
\label{Eq-02}
(\Theta, \Psi) := \hspace{1mm}\biggr(\frac{X}{\mathcal{L}(T)},\hspace{1mm} \frac{Y}{\mathcal{L}(T)}\biggr),\hspace{4mm}\text{where $\mathcal{L}(T) := \mathcal{R}(C)T+1$}\hspace{1mm}.
\end{equation}
Hence, the mechanism for the pattern formation and dispersion on the growing domain is represented, with diffusion coefficients ($D_M$, $D_A$ and $D_C[cm^2/sec]$ for $M$, $A$ and $C$, respectively) and positive constants ($k_1[1/sec]$, $k_2[mL/(cell\cdot sec)]$, $k_3[1/sec]$) as follows:
\begin{eqnarray}
\label{Eq-03}
&& \frac{\partial M}{\partial T} = \frac{D_M}{\mathcal{L}^2(T)}\nabla^2 M + k_1 \hspace{1mm}\chi(A-A_M)\hspace{1mm} M\Biggr(1-\frac{M}{K_M}\Biggr) - \frac{2\mathcal{R}(C)}{\mathcal{L}(T)}M \nonumber \\
&& \nonumber \\
&& \frac{\partial (A-A_{th})}{\partial T} = \frac{D_A}{\mathcal{L}^2(T)}\nabla^2 (A-A_{th}) + k_2 \hspace{1mm}(A-A_{th})\Biggr(1-\frac{A-A_{th}}{K_A}\Biggr) \hspace{1mm} C\Biggr(1-\frac{C}{K_C}\Biggr) - \frac{2\mathcal{R}(C)}{\mathcal{L}(T)}(A-A_{th}) \nonumber \\
&& \nonumber \\
&& \frac{\partial C}{\partial T} = D_C\nabla^2 C + k_3 C\Biggr(1-\frac{C}{K_C}\Biggr)
\end{eqnarray}
where $\nabla^2 = (\partial^2/\partial X^2 + \partial^2/\partial Y^2)$ and a step function, $\chi(u)$, is defined as:
\begin{equation}
\label{Eq-04}
\chi(u) = \left\{
\begin{array}{ll}
1 & (u \geq 0)\\
0 & (u < 0)\hspace{1mm}.
\end{array}
\right.
\end{equation}
When the strain rate becomes approximately equal to $R$, then $C\approx K_C$. And, when $C$ saturates approximately at $K_C$, which is when the effective application of the growing-domain effects begins, the effects for $C$ can be neglected. In addition, when $C$ is away from $K_C$, the strain rate is very small, and they can be also neglected. Furthermore, we are interested in the pattern dispersion after $C$ reaches saturation approximately at $K_C$. Hence, note that the equation for $C$ above does not incorporate the effects.

To nondimensionalize the equation ($\ref{Eq-03}$), the dimensionless quantities below are adopted:
\begin{eqnarray}
\label{Eq-05}
&& m:=\frac{M}{K_M},\hspace{4mm}a:=\frac{A-A_{th}}{K_A},\hspace{4mm}c:=\frac{C}{K_C},\hspace{4mm}t:=k_3 T,\hspace{4mm}\nabla^{*2}:=\frac{D_C}{k_3}\nabla^2 \nonumber \\
&& d_m:=\frac{D_M}{D_C},\hspace{4mm}d_a:=\frac{D_A}{D_C},\hspace{4mm}\sigma:=\frac{k_1}{k_3},\hspace{4mm}\gamma:=\frac{k_2 K_C}{k_3},\hspace{4mm}r:=\frac{R}{k_3},
\end{eqnarray}
which introduces the following nondimensionalized equation:
\begin{eqnarray}
\label{Eq-06}
\frac{\partial m}{\partial t} &=& \frac{d_m}{l^2(t)}\nabla^2 m + \sigma \hspace{1mm}\chi(a-a_m)\hspace{1mm} m(1-m) - \frac{2r(c)}{l(t)}m\nonumber \\
\frac{\partial a}{\partial t} &=& \frac{d_a}{l^2(t)}\nabla^2 a + \gamma \hspace{1mm}a(1-a) \hspace{1mm} c(1-c) - \frac{2r(c)}{l(t)}a \nonumber \\
\frac{\partial c}{\partial t} &=& \nabla^2 c + c(1-c) \nonumber \\
l(t) &:=& r(c)t+1,\hspace{4mm}r(c) := \frac{rh}{h+(c-1)^2}
\end{eqnarray}
with the asterisk for $\nabla^2$ ($=\partial^2 / \partial x^2 + \partial^2 / \partial y^2$) omitted to simplify discussions. Note that $a_m := (A_M-A_{th})/K_A$ based on the dimensionless quantity ($\ref{Eq-05}$).
And hence $a$ is nondimensionalized such that the Dirichlet boundary condition is required as follows:
\begin{equation}
\label{Eq-07}
a\hspace{2mm}\Big|_{\partial\Omega} = 0
\end{equation}
to strictly control the concentration of retinoic acids with the threshold $A_{th}$. Specifically, when the $c$ reaches $1$ (\textit{i.e.}, when the domain growth begins) the equations for $m$ and $a$ in the system ($\ref{Eq-06}$) are described as:
\begin{eqnarray}
\label{Eq-08}
\frac{\partial m}{\partial t} &=& \frac{d_m}{l^2(t)}\nabla^2 m + \sigma \hspace{1mm}\chi(a-a_m)\hspace{1mm} m(1-m) -\frac{2r}{l(t)}m \nonumber \\
\nonumber \\
\frac{\partial a}{\partial t} &=& \frac{d_a}{l^2(t)}\nabla^2 a -\frac{2r}{l(t)}a
\end{eqnarray}
Therefore, the solution to the second equation in the system ($\ref{Eq-08}$) at sufficiently large $t$ compared to $t_g$ is as follows\footnote{see \ref{App-02} for intermediate formulae}, with a Fourier coefficient, $\mathcal{A}_{i,j}$, which is calculated based on an initial condition:
\begin{equation}
\label{Eq-09}
a(t,x,y) = \frac{1}{l^2(t)}\sum_{i,j\in\mathbb{Z}} \mathcal{A}_{i,j}\hspace{1mm}\cos\frac{i\pi x}{L} \hspace{1mm}\cos\frac{j\pi y}{L} \hspace{1mm} exp\biggr( -\frac{d_a\pi^2(i^2+j^2)t}{l(t)L^2} \hspace{1mm}\biggr)
\end{equation}
which suggests that the uniform static state $a\rightarrow 0$ at sufficiently time $t$. And, based on $a\rightarrow 0$, the first equation in the system ($\ref{Eq-08}$) is represented as:
\begin{equation}
\label{Eq-10}
\frac{\partial m}{\partial t} \approx \frac{d_m}{l^2(t)}\nabla^2 m -\frac{2r}{l(t)}m,\hspace{4mm}\text{at sufficiently large $t$}
\end{equation}
which also suggests that the uniform static state $m\rightarrow 0$ at sufficiently time $t$. Mathematically, the analysis reveals that as the domain expands, the previously formed black patterns ($m$) disperse, resulting in the whitening of the plastron ($a$). When comparing this mathematical observation to the actual phenomenon observed in \textit{P.sinensis}, it is suggested that there exists an unknown threshold $A^*_{th} \approx 0$. This threshold indicates that the concentration of retinoic acids may be controlled with a low threshold in adult \textit{P.sinensis}, even under the stricter condition where the precursors are not depleted hypothetically.
\section{Result}
\begin{figure}[H]

\begin{minipage}[b]{0.45\linewidth}
\begin{center}
\includegraphics[scale=0.30]{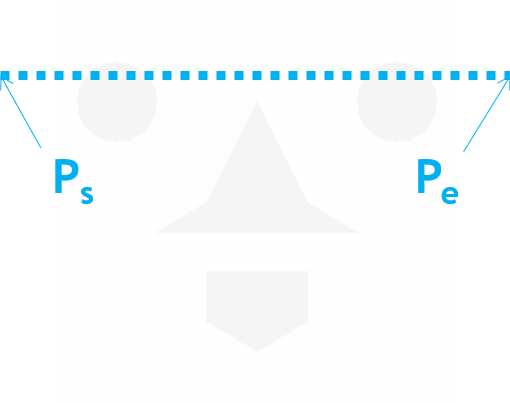}
\subcaption{$m$ on the entire domain at $t=10$ using the equation ($\ref{Eq-06}$), with the intensity of color proportional to the magnitude of $m$}\label{Fig-2_1_all}
\end{center}
\end{minipage}
\begin{minipage}[b]{0.45\linewidth}
\begin{center}
\includegraphics[scale=0.30]{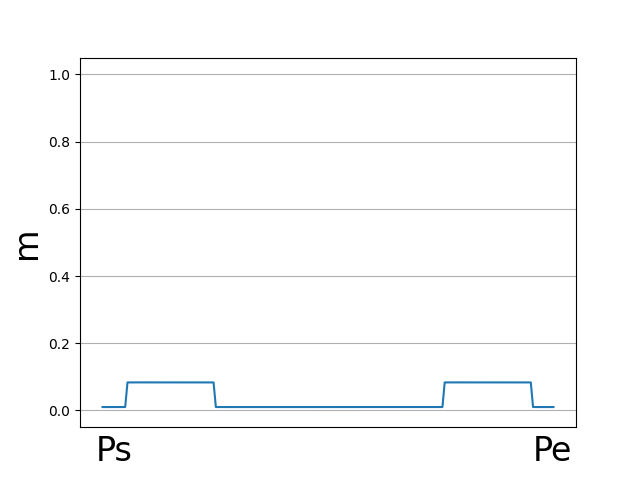}
\subcaption{the concentration gradient of $m$ at $t=10$ using the equation ($\ref{Eq-06}$) along the dashed line shown in Fig.$\ref{Fig-2_1_all}$}\label{Fig-2_1_x45}
\end{center}
\end{minipage}\\

\begin{minipage}[b]{0.45\linewidth}
\begin{center}
\includegraphics[scale=0.30]{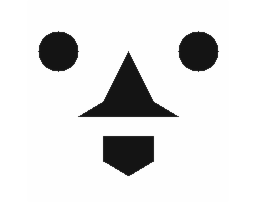}\\
\subcaption{$m$ on the entire domain at $t=100$ using the equation ($\ref{Eq-06}$), with the intensity of color proportional to the magnitude of $m$}\label{Fig-2_2_all}
\end{center}
\end{minipage}
\begin{minipage}[b]{0.45\linewidth}
\begin{center}
\includegraphics[scale=0.30]{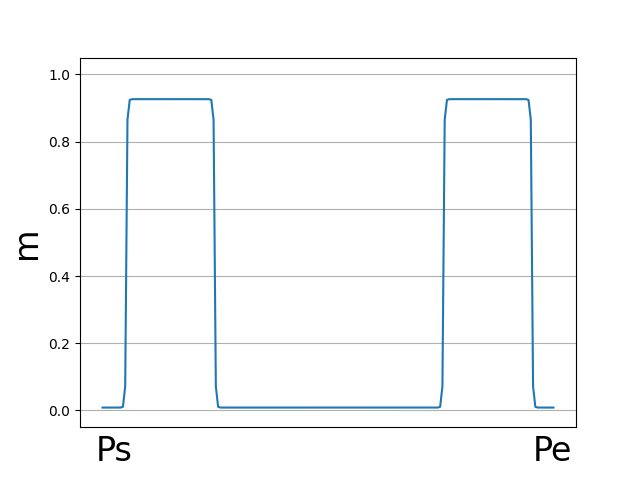}\\
\subcaption{the concentration gradient of $m$ at $t=100$ using the equation ($\ref{Eq-06}$) along the dashed line shown in Fig.$\ref{Fig-2_1_all}$}\label{Fig-2_2_x45}
\end{center}
\end{minipage}\\

\begin{minipage}[b]{0.45\linewidth}
\begin{center}
\includegraphics[scale=0.30]{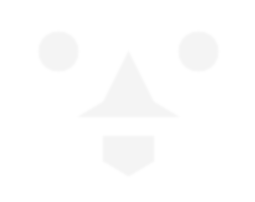}
\subcaption{$m$ on the entire domain at $t=2500$ using the equation ($\ref{Eq-06}$), with the intensity of color proportional to the magnitude of $m$}\label{Fig-2_3_all}
\end{center}
\end{minipage}
\begin{minipage}[b]{0.45\linewidth}
\begin{center}
\includegraphics[scale=0.30]{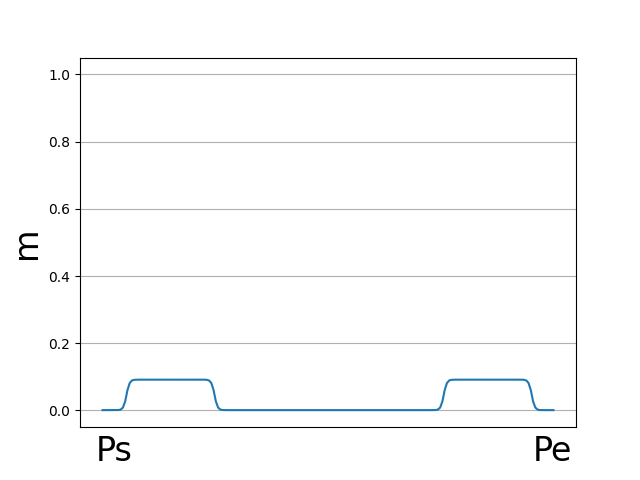}
\subcaption{the concentration gradient of $m$ at $t=2500$ using the equation ($\ref{Eq-06}$) along the dashed line shown in Fig.$\ref{Fig-2_1_all}$}\label{Fig-2_3_x45}
\end{center}
\end{minipage}\\

\begin{minipage}[b]{0.45\linewidth}
\begin{center}
\includegraphics[scale=0.30]{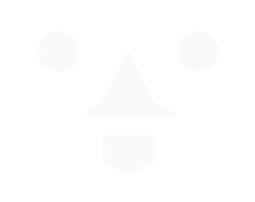}
\subcaption{$m$ on the entire domain at $t=4000$ using the equation ($\ref{Eq-06}$), with the intensity of color proportional to the magnitude of $m$}\label{Fig-2_4_all}
\end{center}
\end{minipage}
\begin{minipage}[b]{0.45\linewidth}
\begin{center}
\includegraphics[scale=0.30]{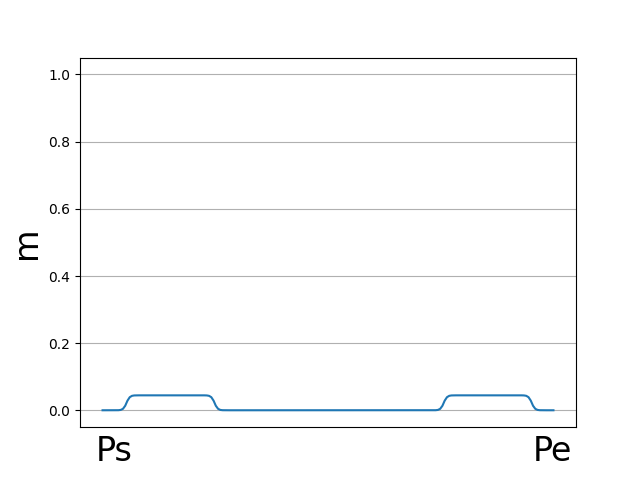}
\subcaption{the concentration gradient of $m$ at $t=4000$ using the equation ($\ref{Eq-06}$) along the dashed line shown in Fig.$\ref{Fig-2_1_all}$}\label{Fig-2_4_x45}
\end{center}
\end{minipage}\\

\caption{Numerical solutions for $m$ at each $t$ using the equation ($\ref{Eq-06}$) under the following parameters: $d_m=d_a=5\times 10^{-4}$, $\sigma=1$, $\gamma=5$, $r=1\times 10^{-3}$, $h=1\times 10^{-4}$ (larger diffusion coefficients used in Fig.$\ref{Fig-c-06}$ in \ref{App-04} for reproductions similar to an actual phenomenon)\label{Fig-2}}
\end{figure}

Mathematically, the solutions based on the equations ($\ref{Eq-09}$) and ($\ref{Eq-10}$) suggest that the black patterns previously formed are dispersed and the vivid orange plastron is whitened, whereas numerical solutions based on the equation ($\ref{Eq-06}$) also shows the phenomenon (the black-pattern transitions on Fig.$\ref{Fig-2}$). The essential point is the dilution terms $-(2r(c)/l(t))a$ and $-(2r(c)/l(t))m$, which were neglected and unexplored, as one of the growing-domain effects. In contrast, another numerical analysis without growing-domain effects on the equation ($\ref{Eq-08}$) below:
\begin{eqnarray}
\label{Eq-11}
\frac{\partial m}{\partial t} &=& d_m\nabla^2 m + \sigma \hspace{1mm}\chi(a-a_m)\hspace{1mm} m(1-m) \nonumber \\
\nonumber \\
\frac{\partial a}{\partial t} &=& d_a\nabla^2 a
\end{eqnarray}
shows that the black patterns are not dispersed (Fig.\ref{Fig-3} in \ref{App-03}). That is because the diffusion of the substances, $a$ and $m$, are much less effective than that of the cell, $c$, as $d_a,d_m\ll 1$ in the nondimensionalized equation ($\ref{Eq-06}$). Similarly a lower strain rate (\textit{e.g.}, $r=1\times10^{-4}$ in Fig.$\ref{Fig-c-02}$ in \ref{App-03}) does not completely disperse the black patterns at a certain $t$ even though the pattern will be dispersed at a sufficiently large $t$, whereas a larger one (\textit{e.g.}, $r=2\times10^{-2}$ in Fig.$\ref{Fig-c-02}$ in \ref{App-03}) does not even form the black patterns, sufficiently. Regarding $\gamma$, on the other hand, when $\gamma$ is relatively smaller then the pattern formation does not even occur, and when $\gamma$ is relatively larger then the pattern is not completely dispersed at a certain $t$ even though the pattern will be dispersed at a sufficiently large $t$ (for instance, in Fig.$\ref{Fig-c-03}$ in \ref{App-03}). This robustness for pattern dispersion is due to the fact that $r(c)$ eventually saturates to $r$ regardless of $t$ (Fig.$\ref{Fig-c-05}$ in \ref{App-03}). What can be emphasized is that the essential point shows pattern-dispersion robustness at the growing-domain stages, even under the stricter conditions.
\\
\indent
Morphologically, one of the distinct points in this study is the tight retinoic-acid control by the enzyme $cyp26b1$ expressed within osteoblasts. In fact, a couple of studies  \cite{bib54, bib63, bib68} have already shown that an excessive concentration of retinols or retinoic acids imposes negative effects on osteogenesis or ossification, which inevitably suggests the tight retinoic-acid control on the plastron of \textit{P. sinensis} in this study which requires the tight boundary condition ($\ref{Eq-07}$) and forms the black patterns induced by the difference of the retinoic-acid concentrations. From another perspective, that may show how essential and preferential osteogenesis and ossification are for a better survival strategy.
\newpage
\section{Discussion}
As a matter of fact, it should be noted that not all black patterns are uniformly dispersed on the plastron of \textit{Pelodiscus} including species other than \textit{P.sinensis}, as depicted in the figures presented in a study \cite{bib50} for instance. This raises the question for future research to identify the factors within the mechanism explored in this study that contribute to variations in pattern formation and dispersion on the plastron. Based on the numerical solutions presented in Fig.$\ref{Fig-c-02}$ and Fig.$\ref{Fig-c-04}$ in \ref{App-03}, it can be observed that the persistence of specific patterns is influenced by both/either a low strain rate ($r$) and/or a high value of $\gamma$.
\\
\indent
On the contrary, there is an additional dark pattern observed on the plastron of certain adult \textit{Pelodiscus} individuals \cite{bib50} that lacks the black patterns mentioned earlier. The skeletal structure of the plastron in \textit{P.sinensis} \cite{bib46} reveals the presence of hyo-, hypo-, and xiphiplastra beneath these dark patterns, suggesting the formation of callosities around them through the callosity mechanism on the horny layer \cite{bib46, bib55} as \textit{P.sinensis} matures.
\\
\indent
In the case of \textit{P.sinensis}, the dorsal part of the carapace does not exhibit a vibrant orange color, even during the embryonic stages. According to a study \cite{bib52}, the main components of the carapace are solely derived from the endoskeleton, indicating the absence of dermal elements known as osteoderms. This suggests that carotenoid pigments may be indirectly regulated by osteoblasts, with the development of costal and neural plates prioritized as a crucial survival strategy for turtles. As for the irregular black-spot patterns observed on the carapace of \textit{P.sinensis}, they could potentially arise from variations in the concentration of keratin bundles on the epidermis \cite{bib49}.
\\
\indent
The figure \cite{bib65} reveals a notable distinction between our findings and the non-uniform orange domain $\Omega^c$ observed in the vicinity of each black pattern during the embryonic stages. While our study mathematically assumes a uniform orange distribution in $\Omega^c$, the non-uniform pattern observed may be attributed to lateral inhibition effects resulting from the logistic increase of retinoic acid concentrations as described in the equation ($\ref{Eq-06}$). Additionally, this study posits that cell $C$ undergoes uniform logistic growth and differentiation. However, it is also plausible to consider that the cell exhibits negative chemotaxis \cite{bib57, bib60}, indirectly interacting with osteoblasts through the enzyme \textit{cyp26b1}. This interaction may induce cell movement towards the regions associated with black patterns, consequently leading to the formation of black patterns due to variations in retinoic acid concentrations.
\section{Conclusion}
This study provides insights into the formation and dispersion of black patterns on the plastron of \textit{Pelodiscus sinensis} and their significance in the species' survival strategy. The black patterns observed on the vivid orange background of the plastron are not intended as warning coloration or for courtship purposes by juveniles, but rather serve as a symbol of the species' survival strategy beyond ecological considerations.
\\
\indent
The mechanism proposed in this study suggests that differences in retinoic acid concentrations on the plastron of \textit{P.sinensis} lead to the formation of black patterns during the embryonic stages. Subsequently, the growing-domain effects, specifically the dilution term, play a crucial role in dispersing the black patterns and whitening the plastron during the growing-domain stages. This mechanism highlights the key role of osteoblasts and the enzyme cyp26b1 in regulating retinoic acid concentrations, with osteogenesis and ossification prioritized as essential elements of the survival strategy. The partial domains where osteogenesis and ossification have not yet occurred need to be thickened for organ system protection with retinoic acid \cite{bib47}, indirectly influencing the concentration of melanin.
\\
\indent
Furthermore, variations in the black patterns observed on the plastron of \textit{P.sinensis} can be attributed to differences in strain rates. The numerical solutions indicate that lower strain rates may not completely disperse the black patterns, leading to a diverse range of patterns on the adult plastron. Additionally, the vivid orange background of the plastron undergoes whitening even under lower strain rates, considering the realistic condition of carotenoid availability.
\\
\indent
The findings of this study have implications for future research in related fields. Understanding the mechanisms underlying pattern formation and dispersion on the plastron of \textit{P.sinensis} can inspire investigations into similar phenomena in other organisms. Moreover, the role of osteoblasts, retinoic acids, and the tight control of their concentrations may have implications beyond pattern formation, providing valuable insights into bone development and regeneration.
\\
\indent
In conclusion, this study not only advances our understanding of pattern formation and dispersion in \textit{P.sinensis} but also highlights the significance of bone biology and retinoic acid regulation in the survival strategy of this species. By delving into the mechanisms underlying pattern formation and dispersion for adaptations, this study paves the way for future research in related fields, fostering a deeper appreciation of the intricate connections between morphology, and ecology.
\section*{Acknowledgements}
This work was supported by funding from Ohagi Hospital, Hashimoto, Wakayama, Japan, and by  Japan Society for the Promotion of Science (JSPS) Topic-Setting Program to Advance Cutting-Edge Humanities and SocialSciences Research Grant Number JPJS00122674991.

\newpage
\renewcommand{\appendix}{}
\appendix
\setcounter{section}{0}
\renewcommand{\thesection}{Appendix.\arabic{section}}
\renewcommand{\thesubsection}{Appendix.\arabic{section}-\arabic{subsection}.}
\section{The reaction-diffusion equation with the growing-domain effects}\label{App-01}
The dimensionless quantities $\bm{x}:=(x,y)$ are defined as $[0,L]\times[0,L]$ with the dimensionless quantity $L$. This study considers a domain such that this two-dimensional domain grows with time. Using the nondimensionalized strain rate $r$ (the dimensionless concentration of cells $c$ is approximately considered to be independent of spatial variables because we require that $c$ be uniformly distributed under an initial condition), the dimensionless distance below is defined as:
\begin{equation}
l(t) := rt +1 \nonumber
\end{equation}
based on which the two-dimensional growing domain is defined as:
\begin{equation}
(\theta, \psi) := \big( (rt+1)x, (rt+1)y \big) \nonumber
\end{equation}
The two-dimensional growing domain is applied to the following reaction-diffusion equation for the concentration $c_s$ of a certain substance $s$:
\begin{equation}
\label{Eq-App-c}
\frac{\partial c_s}{\partial t} = d\Biggr( \frac{\partial^2}{\partial\theta^2}+\frac{\partial^2}{\partial\psi^2} \Biggr)c_s + f(c_s)
\end{equation}
where $d>0$ is constant and $f(c_s)$ is a reaction term. And, the diffusion term with $x$ and $y$ is represented as:
\begin{equation}
\frac{1}{(rt+1)^2} \cdot d\Biggr( \frac{\partial^2}{\partial x^2}+\frac{\partial^2}{\partial y^2} \Biggr)c_s\hspace{1mm}. \nonumber
\end{equation}
Changes in the area $l^2$ and the substance quantity $s$ after $\Delta t$ are defined as:
\begin{eqnarray}
&& (l+\Delta l)^2 \nonumber \\
&& s+\Delta s \nonumber
\end{eqnarray}
which introduces the following change in the concentration $c_s$:
\begin{equation}
\Delta c_s = \frac{s+\Delta s}{(l+\Delta l)^2} - \frac{s}{l^2}\hspace{1mm}. \nonumber
\end{equation}
$\Delta l=r\Delta t$ since $dl/dt=r$, and the change in the substance $s$ depends on the reaction term in the equation ($\ref{Eq-App-c}$),
\begin{equation}
\Delta s = l^2 f(c_s)\Delta t\hspace{1mm}. \nonumber
\end{equation}
Hence,
\begin{equation}
\Delta c_s = \frac{s+l^2 f(c_s)\Delta t}{(l+r\Delta t)^2} -c_s = \frac{f(c_s) -2c_s r/l - c_s r^2\Delta t/l^2}{1+2\Delta l/l + \Delta l^2/l^2} \Delta t\hspace{1mm}. \nonumber
\end{equation}
which leads to the following:
\begin{equation}
\lim_{\Delta t\rightarrow 0} \frac{\Delta c_s}{\Delta t} = f(c_s) -\frac{2r}{rt+1}c_s \nonumber
\end{equation}
where $l=rt+1$ . Thus, the equation ($\ref{Eq-App-c}$) is represented with the fixed $(x,y)$ ($[0,L]\times [0,L]$) as follows:
\begin{eqnarray}
&& \frac{\partial c_s}{\partial t} = D(t)\Biggr( \frac{\partial^2}{\partial x^2}+\frac{\partial^2}{\partial y^2} \Biggr)c_s + F(c_s) \nonumber \\
&& D(t) := \frac{d}{(rt+1)^2} \nonumber \\
&& F(c_s) := f(c_s) -\frac{2r}{rt+1}c_s \hspace{1mm}. \nonumber
\end{eqnarray}
\section{The derivation of the equation (\ref{Eq-09})}\label{App-02}
The equation ($\ref{Eq-09}$) is derived by using the method of variable separation, $a(t,x,y)=:T(t)\cdot X(x)\cdot Y(y)$, as follows:
\begin{eqnarray}
&& XY\frac{dT}{dt} = \frac{d_a}{l^2(t)} \Biggr\{ YT\frac{d^2 X}{dx^2} + XT\frac{d^2 Y}{dy^2} \Biggr\} - \frac{2r}{l(t)}XYT \nonumber \\
&& \nonumber \\
&& \Rightarrow\quad \frac{l^2(t)}{d_a}\biggr( \hspace{1mm}\frac{1}{T}\frac{dT}{dt} + \frac{2r}{l(t)} \biggr) = \frac{1}{X}\frac{d^2 X}{dx^2} + \frac{1}{Y}\frac{d^2 Y}{dy^2} =: -\lambda^2,\hspace{4mm} \text{where}\hspace{2mm} \lambda>0\hspace{1mm}. \nonumber
\end{eqnarray}
Then, $\lambda$ is defined as:
\begin{equation}
\lambda^2 = \lambda_i^2 + \lambda_j^2,\hspace{4mm}\lambda_i,\lambda_j>0,\hspace{4mm}i,j\in\mathbb{Z}\hspace{1mm}, \nonumber
\end{equation}
and the following is the equation to be solved:
\begin{eqnarray}
&& \frac{1}{X}\frac{d^2 X}{dx^2} = -\lambda_i^2 \nonumber \\
&& \nonumber \\
&& \frac{1}{Y}\frac{d^2 Y}{dy^2} = -\lambda_j^2 \nonumber \\
&& \nonumber \\
&& \frac{l^2(t)}{d_a}\biggr( \hspace{1mm}\frac{1}{T}\frac{dT}{dt} + \frac{2r}{l(t)} \biggr) = -(\lambda_i^2 + \lambda_j^2)\hspace{1mm}. \nonumber
\end{eqnarray}
Based on the zero-flux boundary condition, $X$ and $Y$ are solved as follows (note that the integration constants are omitted):
\begin{eqnarray}
&& X(x) = \cos \lambda_i x ,\hspace{4mm}\lambda_i = \frac{i\pi}{L} \nonumber \\
&& \nonumber \\
&& Y(y) = \cos \lambda_j y ,\hspace{4mm}\lambda_j = \frac{j\pi}{L}\hspace{1mm}. \nonumber
\end{eqnarray}
These are substituted into the equation for $T$, and $T$ is solved as follows:
\begin{eqnarray}
&& \int\frac{dT}{T} = \int_{t_g}^t\biggr( -\frac{2r}{l(t)} -\frac{d_a(\lambda_i^2 + \lambda_j^2)}{l^2(t)}\biggr) dt = \biggr[ -2\log l(t) +\frac{d_a(\lambda_i^2 + \lambda_j^2)}{rl(t)} \biggr]_{t_g}^t \nonumber \\
&& \nonumber \\
&& \Rightarrow\quad T(t) = \frac{l^2(t_g)}{l^2(t)}\hspace{1mm}exp\biggr(-\frac{d_a\pi^2(i^2 + j^2)(t-t_g)}{l(t_g)\hspace{1mm}l(t)\hspace{1mm}L^2} \hspace{1mm}\biggr) \nonumber \\
&& \nonumber \\
&& \hspace{15mm} \approx \frac{1}{l^2(t)}\hspace{1mm}exp\biggr(-\frac{d_a\pi^2(i^2 + j^2)t}{l(t)L^2} \hspace{1mm}\biggr),\hspace{4mm}\text{where } t\gg t_g\hspace{1mm}. \nonumber
\end{eqnarray}
\section{The comparative results by numerical solutions of the system (\ref{Eq-11})}\label{App-03}
While the pattern dispersion in the system ($\ref{Eq-06}$) is shown by the numerical analysis which is performed using Python source code that incorporates the growing-domain effects in a general reaction-diffusion system, pattern dispersion is not confirmed by the numerical solutions of the system ($\ref{Eq-11}$) under the same parameters below:
\begin{figure}[H]

\begin{minipage}[b]{0.45\linewidth}
\begin{center}
\includegraphics[scale=0.30]{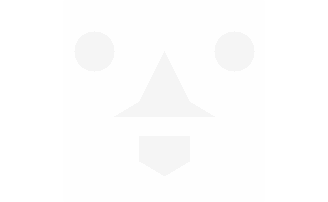}
\subcaption{$m$ on the entire domain at $t=10$ using the equation ($\ref{Eq-11}$), with the intensity of color proportional to the magnitude of $m$}\label{Fig-3_1_all}
\end{center}
\end{minipage}
\begin{minipage}[b]{0.45\linewidth}
\begin{center}
\includegraphics[scale=0.30]{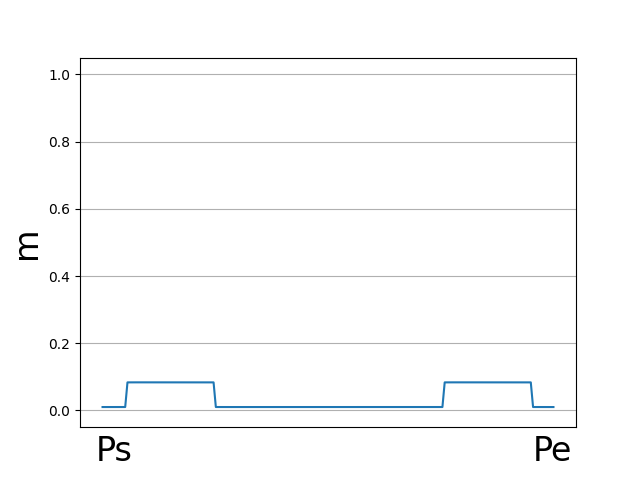}
\subcaption{the concentration gradient of $m$ at $t=10$ using the equation ($\ref{Eq-11}$) along the dashed line shown in Fig.$\ref{Fig-2_1_all}$}\label{Fig-3_1_x45}
\end{center}
\end{minipage}\\

\begin{minipage}[b]{0.45\linewidth}
\begin{center}
\includegraphics[scale=0.30]{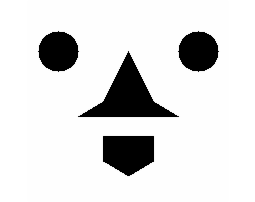}
\subcaption{$m$ on the entire domain at $t=100$ using the equation ($\ref{Eq-11}$), with the intensity of color proportional to the magnitude of $m$}\label{Fig-3_2_all}
\end{center}
\end{minipage}
\begin{minipage}[b]{0.45\linewidth}
\begin{center}
\includegraphics[scale=0.30]{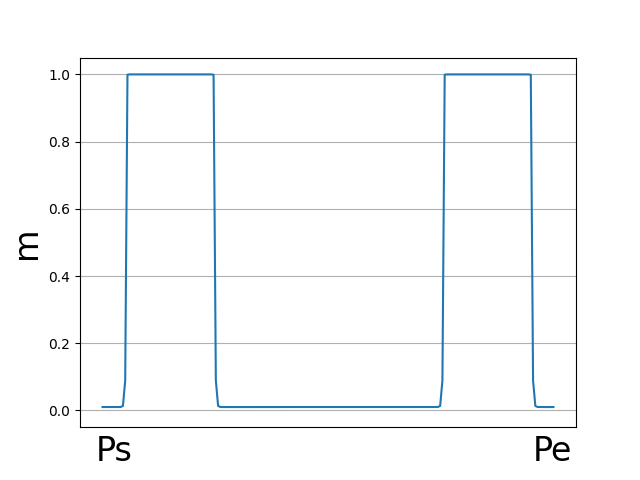}
\subcaption{the concentration gradient of $m$ at $t=100$ using the equation ($\ref{Eq-11}$) along the dashed line shown in Fig.$\ref{Fig-2_1_all}$}\label{Fig-3_2_x45}
\end{center}
\end{minipage}\\

\begin{minipage}[b]{0.45\linewidth}
\begin{center}
\includegraphics[scale=0.30]{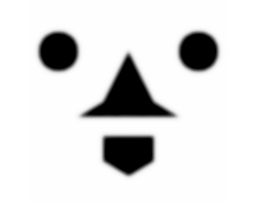}
\subcaption{$m$ on the entire domain at $t=2500$ using the equation ($\ref{Eq-11}$), with the intensity of color proportional to the magnitude of $m$}\label{Fig-3_3_all}
\end{center}
\end{minipage}
\begin{minipage}[b]{0.45\linewidth}
\begin{center}
\includegraphics[scale=0.30]{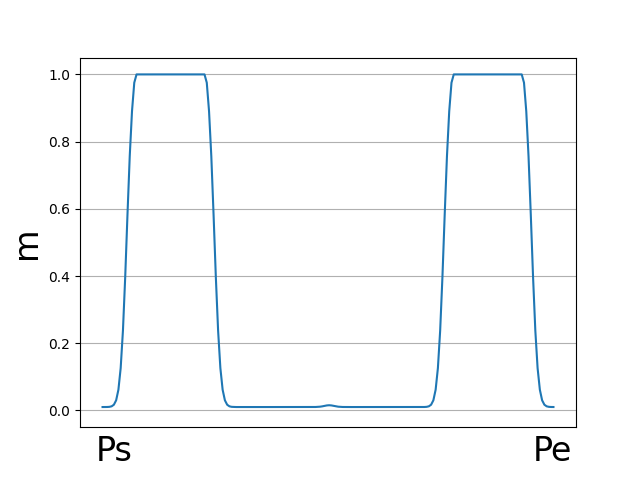}
\subcaption{the concentration gradient of $m$ at $t=2500$ using the equation ($\ref{Eq-11}$) along the dashed line shown in Fig.$\ref{Fig-2_1_all}$}\label{Fig-3_3_x45}
\end{center}
\end{minipage}\\

\begin{minipage}[b]{0.45\linewidth}
\begin{center}
\includegraphics[scale=0.30]{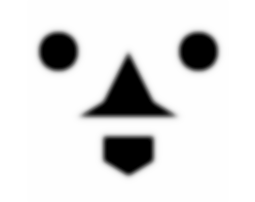}
\subcaption{$m$ on the entire domain at $t=4000$ using the equation ($\ref{Eq-11}$), with the intensity of color proportional to the magnitude of $m$}\label{Fig-3_4_all}
\end{center}
\end{minipage}
\begin{minipage}[b]{0.45\linewidth}
\begin{center}
\includegraphics[scale=0.30]{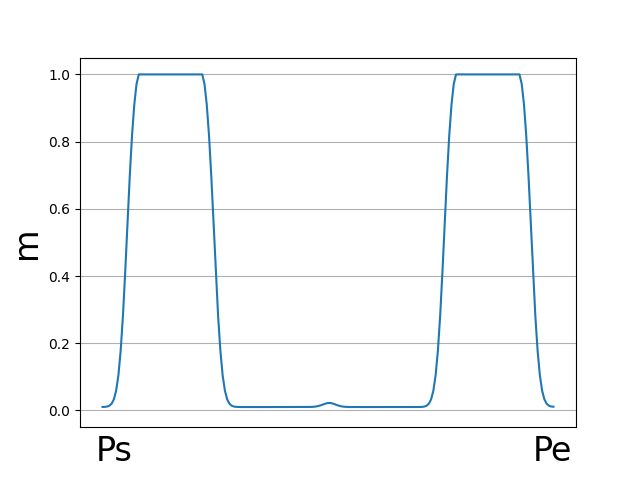}
\subcaption{the concentration gradient of $m$ at $t=4000$ using the equation ($\ref{Eq-11}$) along the dashed line shown in Fig.$\ref{Fig-2_1_all}$}\label{Fig-3_4_x45}
\end{center}
\end{minipage}\\

\caption{Numerical solutions for $m$ at each $t$ using the equation ($\ref{Eq-11}$), \textit{i.e.}, without the Growing-domain effects, under the same parameters: $d_m=d_a=5\times 10^{-4}$, $\sigma=1$, $\gamma=5$, $r=1\times 10^{-3}$, $h=1\times 10^{-4}$\label{Fig-3}}
\end{figure}
\section{The comparison of numerical analysis results by changing the numerical values of each parameter, $r$, $\gamma$, $h$, and diffusion coefficients}\label{App-04}
\begin{figure}[H]
\begin{center}
\includegraphics[scale=0.50]{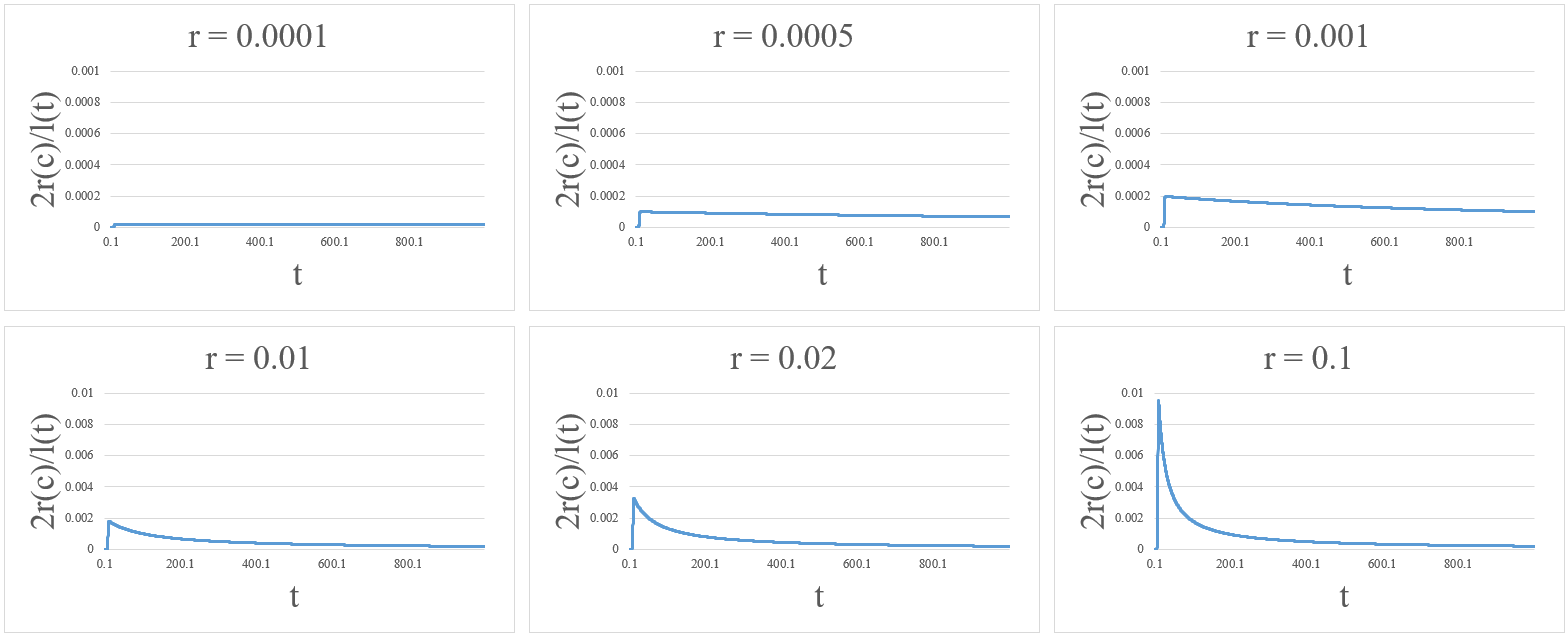}
\end{center}
\caption{Numerical solutions for $2r(c)/l(t)$ at the point $P$ using the equation ($\ref{Eq-06}$) with different values of $r$, under the following parameters and conditions: $\gamma=5$, $\sigma=1$, $h=1\times10^{-4}$, and $t\leq1000$\label{Fig-c-03}}
\end{figure}

At the point $P$ below, numerical analysis will be performed by changing each parameter, $r$, $\gamma$, and $h$, to study how the pattern formation and dispersion will be affected:
\begin{figure}[H]
\begin{center}
\includegraphics[scale=0.30]{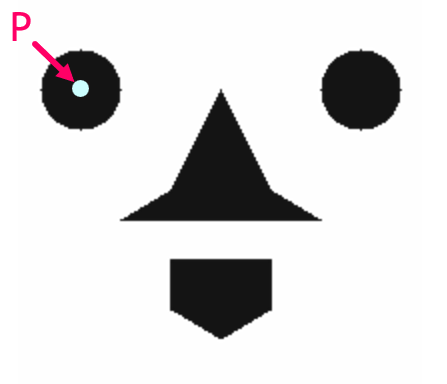}
\end{center}
\caption{The point $P$, $m(50, 30)$, is observed in the following numerical analysis.\label{Fig-c-01}}
\end{figure}
\noindent
As Fig.$\ref{Fig-c-02}$ shows below, when $r$ is relatively smaller, the pattern is formed at the point $P$ but is less likely to disperse at a certain $t$, whereas when $r$ is relatively larger, the pattern is less likely to be formed in the first place:
\begin{figure}[H]
\begin{center}
\includegraphics[scale=0.50]{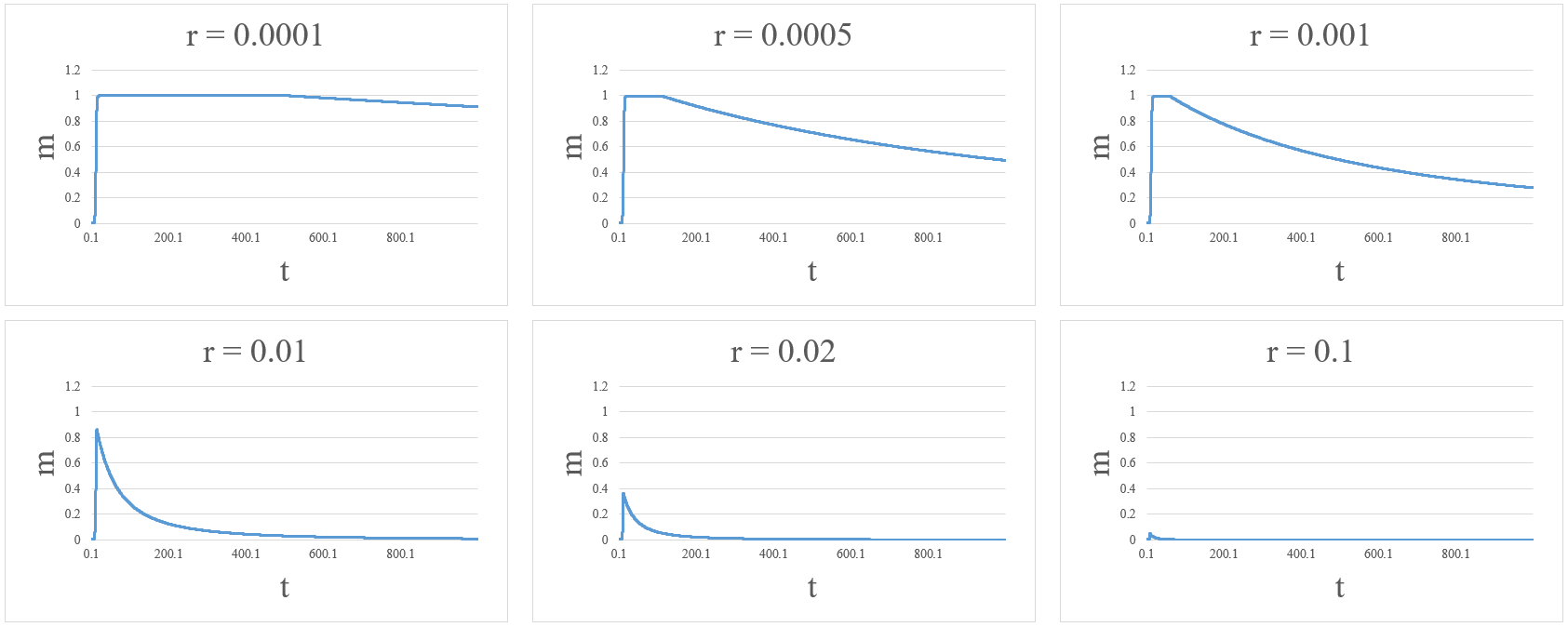}
\end{center}
\caption{Numerical solutions for $m$ at the point $P$ using the equation ($\ref{Eq-06}$) with different values of $r$, under the following parameters and conditions: $\gamma=5$, $\sigma=1$, $h=1\times10^{-4}$, and $t\leq1000$\label{Fig-c-02}}
\end{figure}
\noindent
This is because there is a tendency for the coefficient $2r(c)/l(t)$ in the equation ($\ref{Eq-06}$) to increase as $r$ increases (Fig.$\ref{Fig-c-03}$), which may be consistent with the intuition that as the strain rate in a certain domain getting larger, the dilution of concentration within the domain will accelerate.
\noindent
Fig.$\ref{Fig-c-04}$, on the other hand, shows that a relatively smaller $\gamma$ does not even form the pattern at the point $P$ due to the term, $\chi(a-a_m)$, in the equation ($\ref{Eq-06}$) (\textit{e.g.}, $a_m:=0.5$), whereas a larger one does not completely disperse the pattern at a certain $t$:
\begin{figure}[H]
\begin{center}
\includegraphics[scale=0.50]{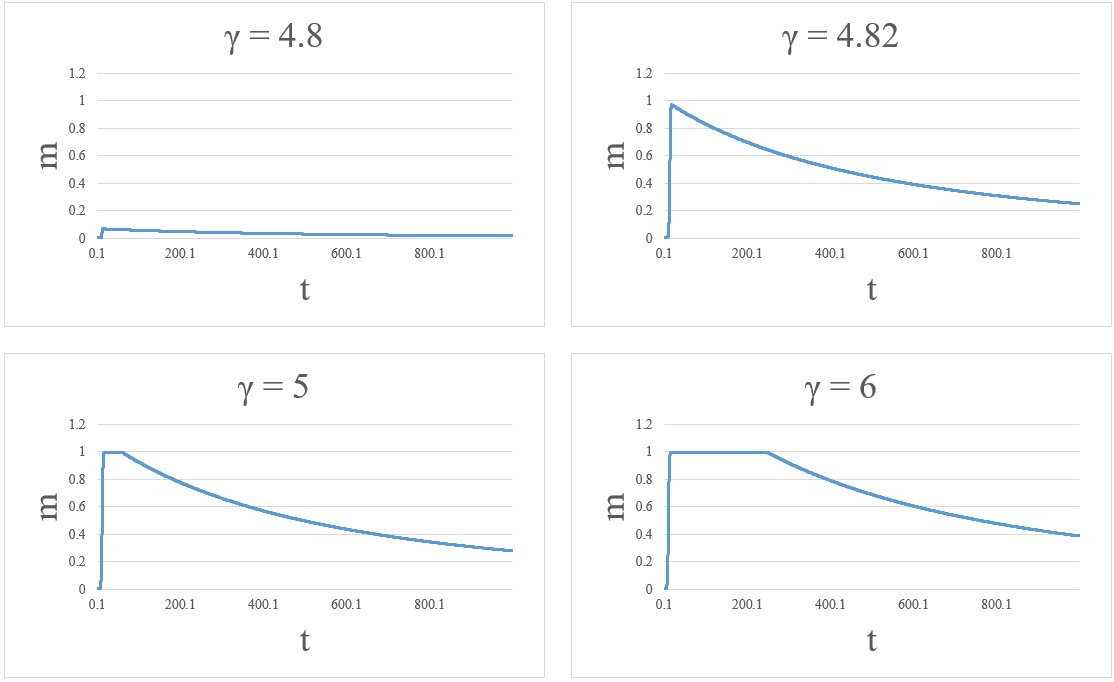}
\end{center}
\caption{Numerical solutions for $m$ at the point $P$ using the equation ($\ref{Eq-06}$) with different values of $\gamma$, under the following parameters and conditions: $\sigma=1$, $r=1\times10^{-3}$, $h=1\times10^{-4}$, and $t\leq1000$\label{Fig-c-04}}
\end{figure}
\noindent
In short, that indicates that both a strain rate ($r(c)$) for a domain and a growth rate ($\gamma$) for $m$ are key factors that affect both pattern formation and dispersion. From a perspective of pattern dispersion, the equation ($\ref{Eq-06}$) shows the robustness for the pattern dispersion at a sufficiently large $t$ which is caused by the function $r(c)$, including the hatching coefficient $h$, saturating to $r$ much earlier than the sufficiently large $t$ below:
\begin{figure}[H]
\begin{center}
\includegraphics[scale=0.50]{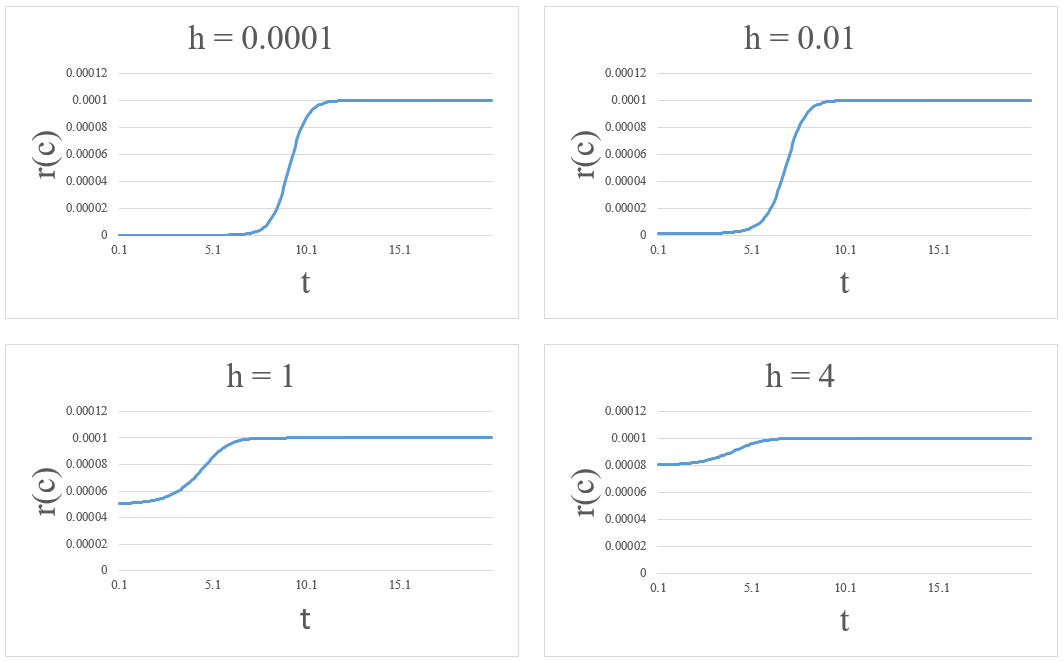}
\end{center}
\caption{Numerical solutions for $r(c)$ at the point $P$ using the equation ($\ref{Eq-06}$) with different values of $h$, under the following parameters and conditions: $\gamma=5$, $\sigma=1$, $r=1\times10^{-3}$, and $t\leq20$\label{Fig-c-05}}
\end{figure}
\noindent
Although the small diffusion coefficients, $d_a=d_m=5\times10^{-4}$, are arbitrarily used so as to enhance the visibility of pattern formation and dispersion for discussions, it would be necessary to use, for instance, the following larger diffusion coefficients than the small ones to obtain pattern formation dispersion similar to an actual phenomenon:
\begin{figure}[H]

\begin{minipage}[b]{0.45\linewidth}
\begin{center}
\includegraphics[scale=0.30]{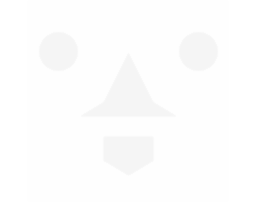}
\subcaption{$m$ on the entire domain at $t=10$ using the equation ($\ref{Eq-06}$), with the intensity of color proportional to the magnitude of $m$}
\end{center}
\end{minipage}
\begin{minipage}[b]{0.45\linewidth}
\begin{center}
\includegraphics[scale=0.30]{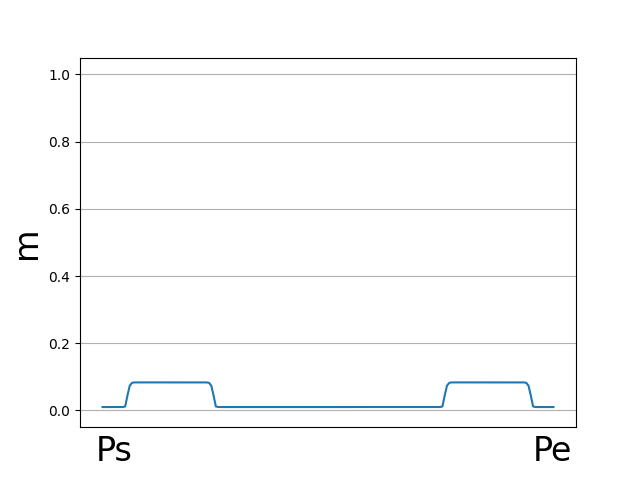}
\subcaption{the concentration gradient of $m$ at $t=10$ using the equation ($\ref{Eq-06}$) along the dashed line shown in Fig.$\ref{Fig-2_1_all}$}
\end{center}
\end{minipage}\\

\begin{minipage}[b]{0.45\linewidth}
\begin{center}
\includegraphics[scale=0.30]{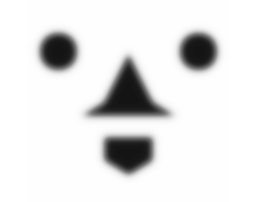}\\
\subcaption{$m$ on the entire domain at $t=100$ using the equation ($\ref{Eq-06}$), with the intensity of color proportional to the magnitude of $m$}
\end{center}
\end{minipage}
\begin{minipage}[b]{0.45\linewidth}
\begin{center}
\includegraphics[scale=0.30]{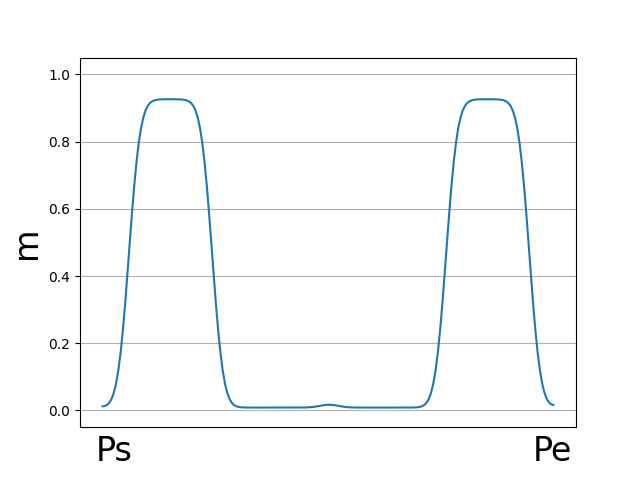}\\
\subcaption{the concentration gradient of $m$ at $t=100$ using the equation ($\ref{Eq-06}$) along the dashed line shown in Fig.$\ref{Fig-2_1_all}$}
\end{center}
\end{minipage}\\

\begin{minipage}[b]{0.45\linewidth}
\begin{center}
\includegraphics[scale=0.30]{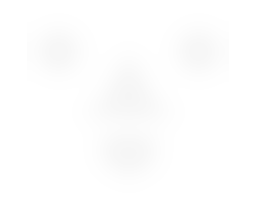}
\subcaption{$m$ on the entire domain at $t=2500$ using the equation ($\ref{Eq-06}$), with the intensity of color proportional to the magnitude of $m$}
\end{center}
\end{minipage}
\begin{minipage}[b]{0.45\linewidth}
\begin{center}
\includegraphics[scale=0.30]{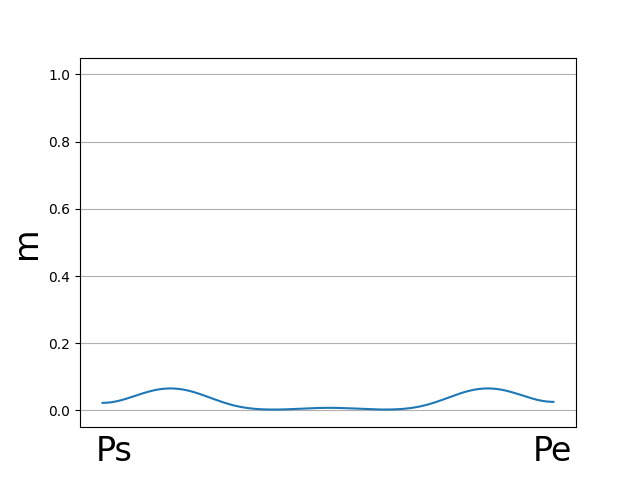}
\subcaption{the concentration gradient of $m$ at $t=2500$ using the equation ($\ref{Eq-06}$) along the dashed line shown in Fig.$\ref{Fig-2_1_all}$}
\end{center}
\end{minipage}\\

\begin{minipage}[b]{0.45\linewidth}
\begin{center}
\includegraphics[scale=0.30]{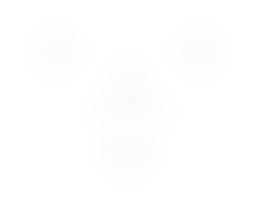}
\subcaption{$m$ on the entire domain at $t=4000$ using the equation ($\ref{Eq-06}$), with the intensity of color proportional to the magnitude of $m$}
\end{center}
\end{minipage}
\begin{minipage}[b]{0.45\linewidth}
\begin{center}
\includegraphics[scale=0.30]{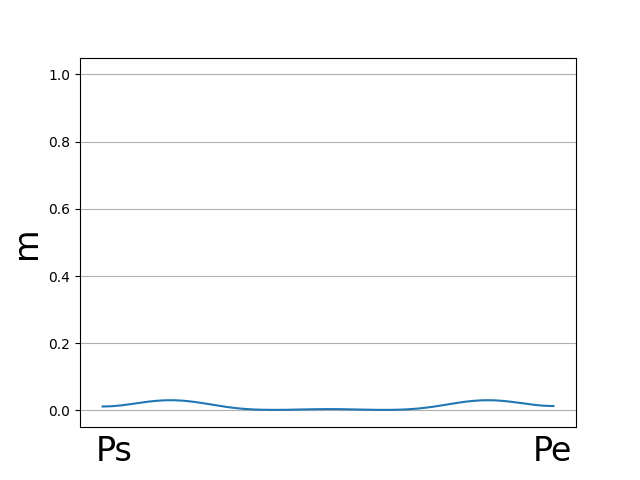}
\subcaption{the concentration gradient of $m$ at $t=4000$ using the equation ($\ref{Eq-06}$) along the dashed line shown in Fig.$\ref{Fig-2_1_all}$}
\end{center}
\end{minipage}\\

\caption{Numerical solutions for $m$ at each $t$ using the equation ($\ref{Eq-06}$) under the following parameters: $d_m=d_a=5\times 10^{-2}$, $\sigma=1$, $\gamma=5$, $r=1\times 10^{-3}$, $h=1\times 10^{-4}$\label{Fig-c-06}}
\end{figure}
\end{document}